\documentclass[prl,aps,reprint,superscriptaddress,floats]{revtex4-1}

\usepackage{amsmath}
\usepackage{epsfig}
\usepackage{color}
\usepackage[bookmarks=true,colorlinks=true,urlcolor=blue,linkcolor=blue,citecolor=blue,breaklinks]{hyperref}
\begin{document}
Published in Phys. Rev. Lett. {\bf 119}, 146101 (2017); DOI: \href{https://doi.org/10.1103/PhysRevLett.119.146101}{10.1103/PhysRevLett.119.146101}

\title{Strong anisotropic interaction controls unusual sticking and scattering of CO at Ru(0001)}
%
%
%

\begin{abstract}
Complete sticking at low incidence energies and broad angular scattering distributions at higher energies are often observed in molecular beam experiments on gas-surface systems which feature a deep chemisorption well and lack early reaction barriers. Although CO binds strongly on Ru(0001), scattering is characterized by rather narrow angular distributions and sticking is incomplete even at low incidence energies. We perform molecular dynamics simulations, accounting for phononic (and electronic) energy loss channels, on a potential energy surface based on first principles electronic structure calculations that reproduce the molecular beam experiments. We demonstrate that the mentioned unusual behavior is a consequence of a very strong rotational anisotropy in the molecule-surface interaction potential. Beyond the interpretation of scattering phenomena, we also discuss implications of our results for the recently proposed role of a precursor state for the desorption and scattering of CO from ruthenium.
\end{abstract}

\author{Ivor Lon\v{c}ari\'c}
\email{ivor.loncaric@gmail.com}
\affiliation{Centro de F\'{\i}sica de Materiales CFM/MPC (CSIC-UPV/EHU), P. Manuel de Lardizabal 5, 20018 Donostia-San Sebasti\'an, Spain}

\author{Gernot F\"{u}chsel}
\affiliation{Leiden Institute of Chemistry, Gorlaeus Laboratories, Leiden University, P.O. Box 9502, 2300 RA Leiden, The Netherlands}
\author{J. I.~Juaristi}
\affiliation{Departamento de F\'{\i}sica de Materiales, Facultad de Qu\'{\i}micas, Universidad del Pa\'{i}s Vasco (UPV/EHU), Apartado 1072, 20080 Donostia-San Sebasti\'an, Spain}
\affiliation{Centro de F\'{\i}sica de Materiales CFM/MPC (CSIC-UPV/EHU), P. Manuel de Lardizabal 5, 20018 Donostia-San Sebasti\'an, Spain}
\affiliation{Donostia International Physics Center DIPC, P. Manuel de Lardizabal 4, 20018 Donostia-San Sebasti\'an, Spain}
\author{Peter Saalfrank}
\affiliation{Institut f\"{u}r Chemie, Universit\"{a}t Potsdam, Karl-Liebknecht-Strasse 24-25, D-14476 Potsdam, Germany}

\maketitle

%
%

Small gas molecules interacting with metal surfaces are of great interest due to their relevance in heterogeneous catalysis. Specifically, the understanding of  the interaction of CO with transition metal surfaces is crucial for two catalytic reactions: the oxidation of CO in car exhaust catalytic converters and the Fischer-Tropsch process. Since ruthenium is the most efficient catalyst for the latter reaction~\cite{Schulz19993}, it is not a surprise that the CO/Ru(0001) system has been the subject of several experimental studies~\cite{PFNUR1980, Starr2013, DellAngela2013, Ostrom2015, bottcher1997, stampfl1997, funk2000, lin2000, ciobica2003, xin2015, beye2013, Gladh201365,over1993,pfnur1983,Buatier1998}. In particular, it has been well established that CO on Ru(0001) binds on a top site vertical to the surface, with the C atom oriented towards the surface~\cite{over1993} and with a large chemisorption energy of E$_{ads}$=1.65 eV~\cite{pfnur1983}.

Molecular beam (MB) experiments have a long standing tradition in surface science and are used to reveal microscopic details about reaction parameters and the potential energy surface (PES)~\cite{Libuda2005157}. Theory is crucial in interpreting experiments by providing information on reactive structures, PES topologies and real-time dynamics.
In this work, we address the dynamics of CO scattering from Ru(0001), for which not yet fully understood MB experiments were performed~\cite{kneitz1999, kneitz1999_2, kleyn2000, menzel1983}. There, high sticking probabilities (close to unity) were determined at low incidence beam energies, and a small dependence of the sticking probability on the surface temperature $T$ (for $T \in[85,385]$~K) was found. These observations are evidence for the presence of a deep chemisorption well directly accessible from vacuum without passing an activation barrier. 
However, experimental results that are still hard to reconcile with the established picture of a highly exothermic, non-activated adsorption process were also obtained: Sticking was incomplete even at very low incidence energies and, at higher incidence energies, the angular-resolved scattering distributions were found to be surprisingly narrow. 
These untypical experimental findings have up to now remained unexplained. 
In what follows, we first briefly discuss why these results were considered as "puzzling" by the experimentalists~\cite{kleyn2000}. 
Subsequently, we present results obtained by molecular dynamics (MD) simulations that reproduce most of the experimental data. Our simulations are carried out on a PES which is based on periodic density functional theory (DFT) calculations, and also incorporate energy dissipation which is crucial in the description of the observed sticking and scattering phenomena.
We will show that the strong rotational anisotropy of the CO/Ru(0001) PES determines the distinct reaction features and that the accurate evaluation of the scattering process on a microscopic level renders the concept of a precursor state proposed in Ref.~\cite{DellAngela2013} moot for the chemisorption process.
Hence, our work demonstrates the importance of state-of-the-art electronic structure methods and converged, multidimensional dissipative dynamics simulations for the understanding of reactive and non-reactive scattering on catalytic surfaces.

The simplest, but often relatively accurate, model to obtain the translational energy loss of a molecule colliding with a surface is the modified Baule model~\cite{Gross2009} treating the collision in the hard sphere limit. 
In its modified version the kinetic energy added to the molecule moving down 
the adsorption well is also taken into account. 
For a molecule with an incidence energy $E_i$ approaching a surface 
characterized by a temperature $T$, this model predicts the energy 
loss $\Delta$ by
\begin{equation}\label{eq:baule}
\Delta=\frac{4\mu}{(1+\mu)^2}\left(E_i+E_\mathrm{ads}-\frac{1}{2}k_\mathrm{B}T\right),
\end{equation}
where $E_\mathrm{ads}$ is the adsorption energy of the molecule on the surface, and $\mu$ is the ratio between the mass of the molecule and the (effective) mass of the surface. 
For the CO/Ru(0001) system, we obtain $\Delta=1.2$~eV by setting the parameters to $E_i=0.1$~eV, $T=300$~K, and $\mu=m_{CO}/m_{Ru}=0.28$. 
Apparently at these conditions the molecule does not have enough energy to escape from the surface after collision. In fact, the Baule model Eq.~\eqref{eq:baule} even suggests, by the condition $\Delta \leq E_i$, complete sticking for $E_i$ up to $3.5$~eV. It is therefore unexpected that at the very low incidence energy of $E_i=0.09$~eV (and $T<300$~K) around 5\% of the CO molecules were observed to scatter from the Ru(0001) surface~\cite{kneitz1999, kneitz1999_2, kleyn2000}. 
High sticking probabilities, but below unity, at low $E_i$ were also measured for CO scattering from other transition metal surfaces such as from Ir(110)~\cite{steinruck1987}, Ni(100)~\cite{DEVELYN1987}, Ni(111)~\cite{tang1986}, Pt(111)~\cite{CAMPBELL1981}, and Pd(111)~\cite{engel1978}. These observations either suggest the complete breakdown of the Baule model or, rather than this, they indicate the presence of an alternative mechanism that prevents CO from reaching the chemisorption well.

Also surprising are the relatively narrow scattering angle distributions [with a full width at half maximum (FWHM) of $22^\circ$] experimentally determined for the CO/Ru(0001) system~\cite{kleyn2000}. Indeed, much broader angular distributions are expected for a system featuring a deep potential well~\cite{kleyn2003}. Scattering distributions are broad if molecules are first deflected toward the surface where they then scatter from locally repulsive walls and bounce to somewhat arbitrary directions. Some
such systems are discussed in Ref.~\cite{Kleyn_2003a}. A particularly interesting example is the scattering of NO from Ru(0001) leading to a scattering distribution with a FWHM of 37$^\circ$, see Ref.~\cite{Butler1997}. Keeping in mind that NO and CO have similar adsorption energies on Ru(0001), the observed narrow width of the angular distribution found for the latter appears even more unusual.

With the aim of understanding these experimental results, we perform MD simulations on a recently developed  PES~\cite{gernot2014,CO_FL}, which accounts for all six degrees of freedom of the molecule relative to the surface.
The PES is obtained using the corrugation reducing procedure~\cite{CRP} to interpolate around $90\,000$ DFT energy points calculated for different configurations of CO on Ru(0001).
In this way, we have constructed an accurate PES, which can efficiently be used in large scale trajectory calculations.
Details on the PES are reported in Refs.~\cite{gernot2014,CO_FL}, but we mention that DFT calculations were carried out using the RPBE functional~\cite{RPBE} with D2~\cite{D2} van der Waals correction, which yields $E_{ads}=1.69$~eV, close to the experimental value.
In our calculations, we also allow for the possibility of energy transfer from the molecule to the surface via (i) excitation of surface atom motion and (ii) electron-hole pair excitations at the surface region.
This is achieved by using the same theoretical approach as outlined in Refs.~\cite{PhysRevLett108096101, MartinGondre2012, ludovic2013, ivor_laser, ivor_laser2, CO_FL}. 
Surface motion as well as surface temperature effects are described by using the generalized Langevin oscillator model (GLO)~\cite{glo3, glo1, glo2}. 
Energy dissipation due to electron-hole pair excitations is modeled by the local density friction approximation~\cite{inaki2008}. 
We note that for the results presented below the effect of electron-hole pair excitations is small while the molecule-phonon coupling is crucial. 
All GLO parameters are given in Ref.~\cite{CO_FL} where we have successfully modeled the laser induced desorption of CO from Ru(0001).
In line with our previous work~\cite{ivor1,gernot2013}, the initial conditions other than the incidence energy and the angle of incidence (such as the orientation and the lateral position of the molecule) are sampled by a conventional Monte-Carlo procedure. For each incidence energy, we evaluate $5000$ trajectories to obtain adsorption probabilities and $500\,000$ trajectories to obtain scattering distributions. Trajectories are propagated up to 12 ps.

\begin{figure}[htb]
\includegraphics[width=\columnwidth]{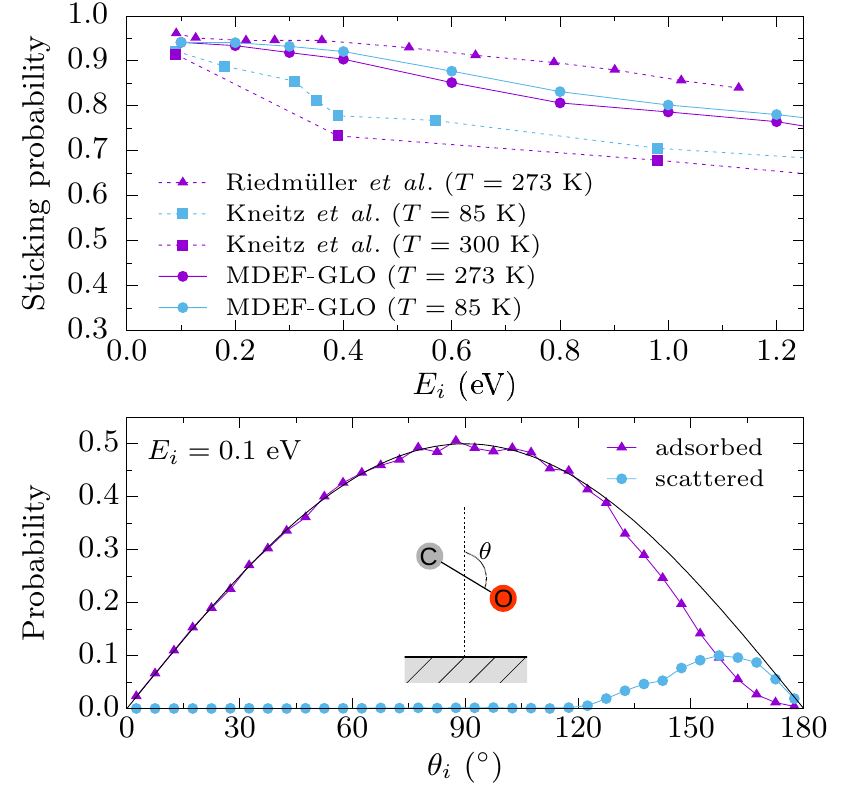}
\caption{(Color online) Upper panel: Calculated sticking probabilities at normal incidence (solid lines and circles) compared to the experimental results (dashed lines) of Kneitz \textit{et al.}~\cite{kneitz1999} (squares) and Riedm\"{u}ller \textit{et al.}~\cite{kleyn2000} (triangles). 
Lower panel: Initial orientation $\theta_i$ of scattered (blue circles) and adsorbed (purple triangles) molecules for $E_i=$~0.1 eV and $T=273$~K. The black line represents the initial distribution of orientations given by $\frac{1}{2}\sin(\theta_i)$.} \label{fig:sticking}
\end{figure}

The upper panel in Fig.~\ref{fig:sticking} compares our computed sticking probability at normal incidence as a function of $E_i$ to corresponding molecular beam experiments of Kneitz \textit{et al.}~\cite{kneitz1999} and Riedm\"{u}ller \textit{et al.}~\cite{kleyn2000}.
We achieve quantitative agreement with experiments at low $E_i$, indicating that our model is able to reproduce the relevant physics. 
At higher $E_i$ there are some differences between
both experiments in the measured sticking coefficients. Since similar experimental set-ups were employed, the reason for the differences is not apparent to us. It is therefore satisfying that our results appear intermediate between the experimental curves. Note also the small dependence of the sticking probability on the surface temperature that is well reproduced by our calculations.

For a given initial energy and angle of incidence, we find that the initial orientation $\theta_i$ of the CO molecule relative to the surface is the most important initial condition to determine the outcome of the dynamics.
We define $\theta_i$ as the angle between the molecular axis and the surface normal with $\theta_i=0^\circ$ ($\theta_i=180^\circ$) describing the molecular orientation at which the Carbon (Oxygen) atom points toward the surface, as depicted in the lower panel of Fig.~\ref{fig:sticking}. Also shown in Fig.~\ref{fig:sticking} is the $\theta_i$ distribution for adsorbed and scattered trajectories computed for $E_i=0.1$~eV. 
We conclude that molecules stick completely to the surface at values $\theta_i<120^\circ$. 
Where we observe complete sticking, the sticking probability curve matches a $\frac{1}{2} \sin(\theta_i)$ function (black line) which represents the initial distribution of $\theta_i$ in the incident beam. However, at values $\theta_i>120^\circ$ the probability for scattering increases and it is close to one for $\theta_i$ near $180^\circ$. This manifests the inability of O-down oriented molecules to reorient and to accommodate in the adsorption well where the C atom points to the surface leading to the observed diminished absolute sticking coefficient. As shown in the figure, due to geometrical reasons the number of molecules in the incident beam with this kind of orientations is anyway very small. On the other hand, at low energies, molecules assuming most probable orientations near $\theta_i\approx 90^\circ$ are able to efficiently reorient and to finally stick to the surface. These facts  explain the observed large, but incomplete, sticking probabilities at low energies.

\begin{figure}[htb]
\includegraphics[width=\columnwidth]{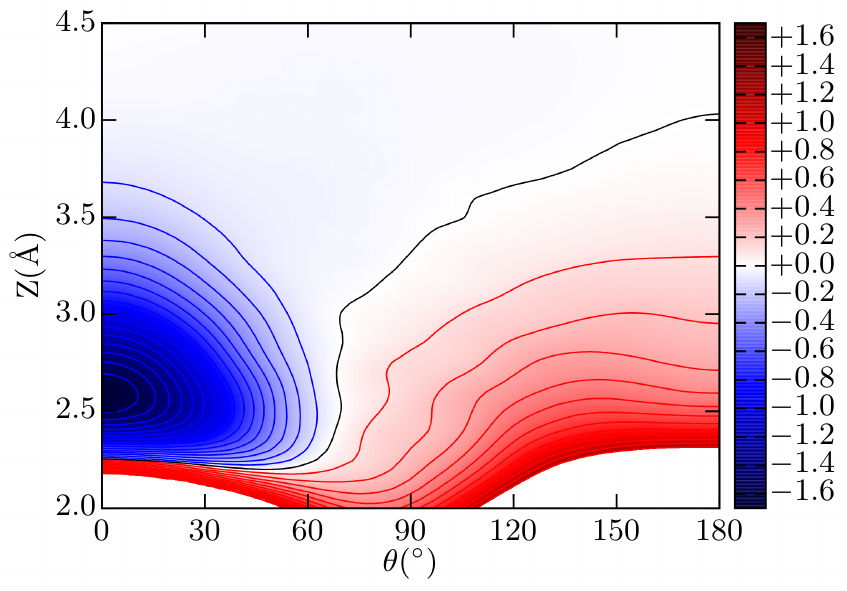}
\caption{(Color online) The color map shows the PES with the center of mass of CO placed at the top site as a function of the molecule-surface distance $Z$ and the orientation angle $\theta$. The other degrees of freedom are optimized. The contour lines are separated by 0.1 eV (blue for negative potential values, black for zero, and red for positive values).} \label{fig:2Dpes}
\end{figure}

The results presented above can be rationalized by Fig.~\ref{fig:2Dpes} where a two dimensional (2D) cut of the PES as a function of the molecular orientation $\theta$ and the molecule-surface distance $Z$ is shown.
For C-down orientations, \textit{i.e.} small values of $\theta$, there appears a deep chemisorption well located at $Z\approx2.5$~\AA. With increasing angle $\theta$, however, the PES becomes more and more repulsive. 
While molecules approaching the surface with small values of $\theta$ are directly attracted to the chemisorption well, their sticking will depend on their ability to lose enough translational energy. 
That is easily achievable at low $E_i$, as discussed above.
A similar outcome is obtained for molecules initially aligned parallel to the surface ($\theta_i\approx90^\circ$). As can be seen in the 2D plot of the PES, they experience considerable reorientation forces that push CO toward the adsorption minimum. 
However, the situation for initially O-down oriented molecules ($\theta_i\approx180^\circ$) is different. For $\theta > 150^\circ$, the repulsive part of the potential is hardly affected by changes of the orientation angle. The resulting reorientation forces are small and molecules scatter back to the gas phase, even at low values of $E_i$.
This demonstrates the importance of the rotational anisotropy of  the interaction potential on sticking at different impact orientations of CO and explains the observed diminished sticking coefficients at low incidence energies.

\begin{figure}[htb]
\includegraphics[width=\columnwidth]{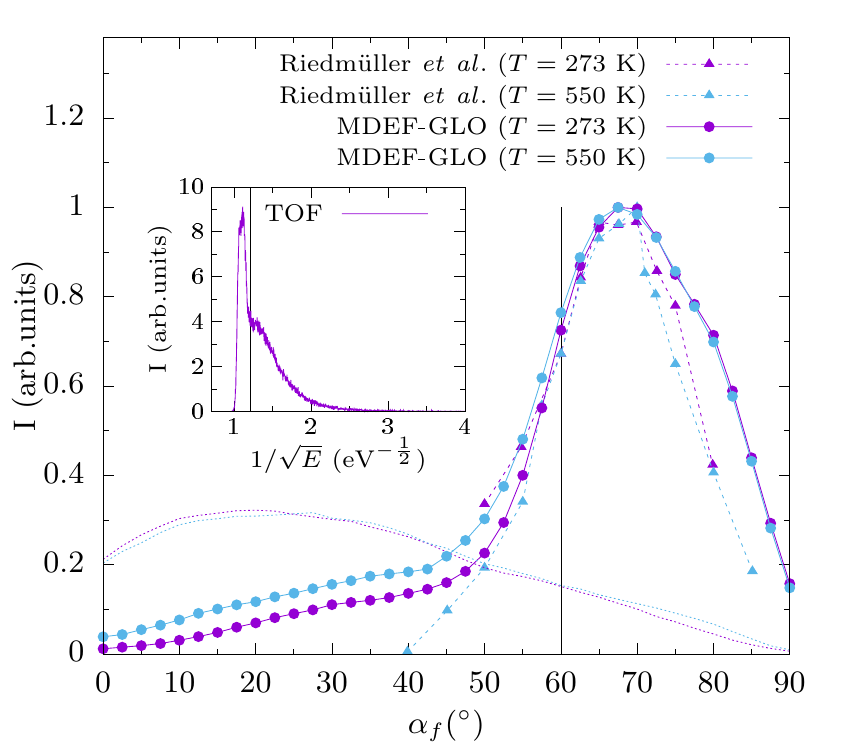}
\caption{(Color online) Angular distributions of \emph{directly scattered molecules} for an incidence angle of $\alpha_i=60^\circ$ (marked by a vertical black line). Our results (solid line with circles) are compared with experiments of Riedm\"uller \textit{et al.}~\cite{kleyn2000} (dashed lines with triangles). Distributions corresponding to \emph{trapped molecules} are shown with dotted lines. 
Inset: Inverse square root of the translational energy of the scattered molecules that is proportional to the time of flight (TOF) of the molecule to the detector. The vertical line separates \emph{directly scattered molecules} (left peak) from \emph{trapped molecules} (right peak) set at $1/\sqrt{E}=1.2~\mathrm{eV}^{-1/2}$.} \label{fig:angular}
\end{figure}

In the following, we show how the rotational anisotropy also controls the narrow width of the scattering angle distributions, the second unusual feature of the CO/Ru(0001) system. 
In Fig.~\ref{fig:angular}, we compare  scattering distributions obtained from our simulations with the experiments of Riedm\"{u}ller \textit{et al.}~\cite{kleyn2000}.
The experiments were performed for CO impinging with $E_i=0.8$~eV in the [11\={2}0] plane of incidence at an angle of incidence of $\alpha_i=60^\circ$. 
The same conditions are applied to our simulations. We used the procedure presented in Ref.~\cite{ludovic2013} to extract molecules that scatter in the plane of incidence. 
In the experiments, only the so-called \emph{directly scattered molecules} were analyzed while the so-called \emph{trapped molecules} were not considered. 
The two types of molecules were distinguished by two separate peaks in the time of flight (TOF) spectra. The left peak (shorter flight time - higher translational energy) was assigned to directly scattered molecules and the right peak (longer flight time - lower translational energy) to trapped molecules. 
As shown in the inset of Fig.~\ref{fig:angular}, our simulations also yield two peaks in the TOF spectrum. Since the experimental procedure used to differentiate between the molecules in the two peaks was not specified, we imply a cut-off as indicated by the vertical line in the inset of Fig.~\ref{fig:angular}, to distinguish both types of molecules.

In Fig.~\ref{fig:angular}, experimental and calculated angular scattering distributions for directly scattered molecules are compared. In accordance with the experiments, we observe a narrow superspecular peak at $\alpha_f\approx70^\circ$ ($\alpha_f$ is the scattering angle measured from the surface normal) and only small effects of the surface temperature on the computed scattering distribution, here characterized by a FWHM of 27$^\circ$. 
As also depicted in the figure, trapped molecules scatter with a broad angular distribution which is expected for a system with a deep chemisorption well. 
We find that molecules initially O-down oriented contribute to the narrow peak, whereas molecules with initial C-down and parallel orientations contribute to the broad peak. 
The experimental observations made on scattering can be explained as follows. 
A narrow scattering angular distribution is caused by early bounced molecules with initial O-down orientation which do not probe the corrugated chemisorption region of the PES.  For the considered scattering conditions, the CO/Ru(0001) potential can be classified as being flat for parallel motions of CO. But it is the flatness of the PES for O-down orientations that keeps the corrugation low while CO can execute a concerted parallel + reorientation motion, see Fig.~\ref{fig:2Dpes}. This is different to C-down (trapped) molecules which form broad distributions. The decorrugation effect on the concerted motion might also explain the narrow scattering distributions found for other CO+metal systems and its absence may lead to the observed broad distributions in the case of NO which also features an anisotropic interaction~\cite{Lahaye_NOPt111}.
The small shift to the superspecular direction comes from the fact that O-down molecules mainly transfer the Z-component of their momentum to the surface.

\begin{figure}[htb]
\includegraphics[width=\columnwidth]{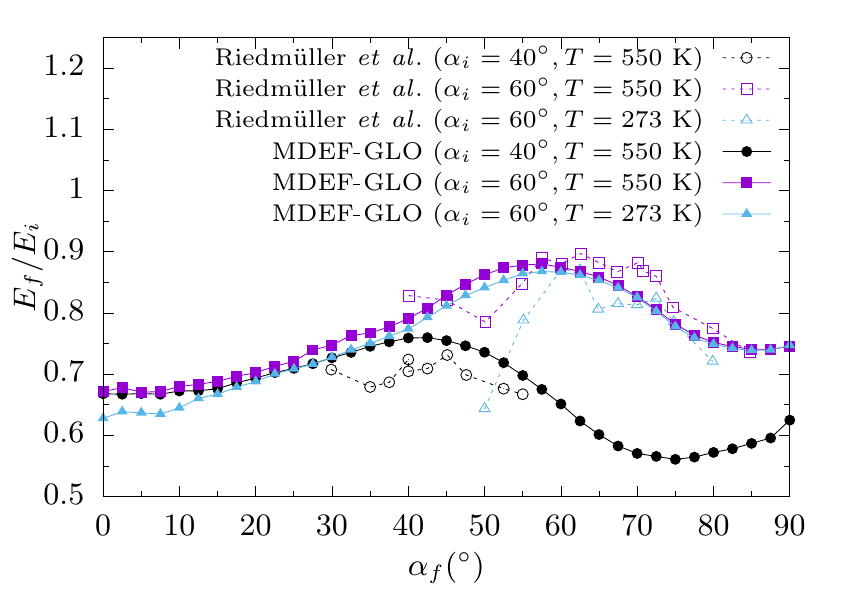}
\caption{(Color online) Dependence of the translational energy loss (shown as the ratio between the initial and final translational energy of the molecule) on the scattering angle. Our results are shown with solid lines and symbols and the experimental results of Riedm\"{u}ller \textit{et al.}~\cite{kleyn2000} are shown with dashed lines and empty symbols. Only directly scattered molecules are taken into account in the analysis.} \label{fig:eloss}
\end{figure}

Riedm\"{u}ller \textit{et al.}~\citep{kleyn2000} also measured the energy loss of directly scattered molecules as a function of the scattering angle. 
For completeness, in Fig.~\ref{fig:eloss} we provide a comparison between the measured and calculated results on  angular-resolved ratios of final to initial translational energy obtained at two different incidence angles $\alpha_i$. 
Our results are in quantitative agreement with experiments, except for two points measured at $T=273$ K and small angles $\alpha_f$ of 50$^\circ$ (55$^\circ$). There, experimental scattering intensities are small (see Fig.~\ref{fig:angular}) which can increase the experimental uncertainty. The achieved good agreement between theory and experiment highlights the credibility of the applied methodology and approximations. This is remarkable in view of
recent advancements representing a big step forward in the description of surface atom motion~\cite{n2ru0001nn,hclaunn}.

This gives us now the opportunity to also comment
on recent femtosecond laser excitation / X-ray detection experiments performed on the system~\cite{DellAngela2013,beye2013}.
There, it was suggested that transient CO dwell in an outer physisorbed
precursor state  separated from a chemisorption state 
due to the occurrence of an entropy barrier.
In appendix B of our previous work~\cite{CO_FL}, 
we provide arguments which rule out its proposed nature, \emph{i.e.}, the entropy barrier.
Although discussed in other CO scattering experiments~\cite{COPt111_Liu}, 
we here do not find indications for a precursor state on 
Ru(0001) since
both, trapped and directly scattered molecules, exhibit similarly short contact times with the surface. 
For trapped molecules, however, the energy loss is larger explaining the TOF data presented above. 
It seems therefore unlikely that a precursor state exists under the assumed 
ideal conditions (zero-coverage, perfectly flat surface). 
Other precursor mechanisms such as molecule-molecule interactions 
inherently present at finite coverages might need to be adduced 
to explain the observed desorption phenomena.
The different preconditions existing in desorption and scattering experiments also question the subtly notion put forward in Ref.~\cite{DellAngela2013} that observations on precursor states made in one experiment warrant conclusions about their presence and nature in the other.

In summary, by performing molecular dynamics simulations with a six-dimensional PES based on DFT calculations, we have been able to quantitatively reproduce experimental results on sticking and scattering for the CO/Ru(0001) system that have previously been regarded as puzzling~\cite{kleyn2000}. 
As such were considered the measured narrow angular distributions of the scattered molecules, and the large but incomplete sticking probabilities 
measured at low incidence energies. 
Our study provides a microscopical elucidation of these results and shows that all experimental observations can be rationalized in terms of the strong rotational anisotropy appearing in the CO/Ru(0001) interaction potential. 
Our work demonstrates that in general multidimensional, statistically converged dynamics performed on an accurate PES and a realistic treatment of energy loss channels are needed to unravel important details about experimental gas-surface scattering data.

I. L. and J. I. J. acknowledge financial support by the Gobierno Vasco-UPV/EHU project IT756-13, and the Spanish Ministerio de Econom{\'\i}a y Competitividad (Grants No. FIS2013-48286-C02-02-P and FIS2016-76471-P). P. S. acknowledges support by Deutsche Forschungsgemeinschaft through project Sa 547/8. G. F. thanks the Nederlandse Organisatie voor Wetenschappelijk Onderzoek (NWO-CW) for financial support through a TOP grant.

\end{document}